\documentclass[%
 reprint,
 superscriptaddress,
 amsmath,amssymb,
 aps,
]{revtex4-1}

\usepackage{amsfonts}
\usepackage{amsmath}
\usepackage{amssymb}
\usepackage{graphicx}
\usepackage{bm}

\begin{document}


\title{ A comprehensive scenario of the single crystal growth and doping dependence of resistivity and anisotropic upper critical fields in (Ba$_{1-x}$K$_x$)Fe$_2$As$_2$ ($0.22 \leq x\leq 1$)}


\author{Y.~Liu}
\email[Corresponding author: ]{yliu@ameslab.gov}
\affiliation{Division of Materials Sciences and Engineering, Ames Laboratory, Ames, Iowa 50011, USA}

\author{M.~A.~Tanatar}
\email[Corresponding author: ]{tanatar@ameslab.gov}
\affiliation{Division of Materials Sciences and Engineering, Ames Laboratory, Ames, Iowa 50011, USA}
\affiliation{Department of Physics and Astronomy, Iowa State University, Ames, Iowa 50011, USA}

\author{W. E. Straszheim}
\affiliation{Division of Materials Sciences and Engineering, Ames Laboratory, Ames, Iowa 50011, USA}
\affiliation{Materials Analysis and Research Laboratory, Iowa State University, Ames, Iowa 50011, USA}

\author{ B.~Jensen }
\affiliation{Division of Materials Sciences and Engineering, Ames Laboratory, Ames, Iowa 50011, USA}

\author{ K.~W.~Dennis }
\affiliation{Division of Materials Sciences and Engineering, Ames Laboratory, Ames, Iowa 50011, USA}

\author{ R.~W.~McCallum }
\affiliation{Division of Materials Sciences and Engineering, Ames Laboratory, Ames, Iowa 50011, USA}
\affiliation{Department of Materials Science and Engineering, Iowa State University, Ames, Iowa 50011, USA}

\author{V. G. Kogan}
\affiliation{Division of Materials Sciences and Engineering, Ames Laboratory, Ames, Iowa 50011, USA}
\affiliation{Department of Physics and Astronomy, Iowa State University, Ames, Iowa 50011, USA}

\author{R.~Prozorov}
\affiliation{Division of Materials Sciences and Engineering, Ames Laboratory, Ames, Iowa 50011, USA}
\affiliation{Department of Physics and Astronomy, Iowa State University, Ames, Iowa 50011, USA}

\author{T.~A.~Lograsso}
\affiliation{Division of Materials Sciences and Engineering, Ames Laboratory, Ames, Iowa 50011, USA}
\affiliation{Department of Materials Science and Engineering, Iowa State University, Ames, Iowa 50011, USA}

\date{\today}


\begin{abstract}

Large high-quality single crystals of hole-doped iron-based superconductor (Ba$_{1-x}$K$_x$)Fe$_2$As$_2$ were grown over a broad composition range $0.22 \leq x \leq 1$ by inverted temperature gradient method.  We found that high soaking temperature, fast cooling rate, and adjusted temperature window of the growth are necessary to obtain single crystals of heavily K doped crystals (0.65$\leq  x \leq$ 0.92) with narrow compositional distributions as revealed by sharp superconducting transitions in magnetization measurements and close to 100\% superconducting volume fraction. The crystals were extensively characterized by x-ray and compositional analysis, revealing monotonic evolution of the $c$-axis crystal lattice parameter with K substitution.
Quantitative measurements of the temperature-dependent in-plane resistivity, $\rho(T)$ found doping-independent, constant within error bars, resistivity at room temperature, $\rho(300K)$, in sharp contrast with significant doping dependence in electron and isovalent substituted BaFe$_2$As$_2$ based compositions. The shape of the temperature dependent resistivity, $\rho(T)$, shows systematic doping-evolution, being close to $T^2$ in overdoped and revealing significant contribution of the $T$-linear component at optimum doping. The slope of the upper critical field, $d H_{c2}/dT$, scales linearly with $T_c$ for both $H\parallel c$, $ H_{c2,c}$, and $H \parallel ab$, $H_{c2,ab}$. The anisotropy of the upper critical field, $\gamma \equiv H_{c2,ab} / H_{c2,c}$ determined near zero-field $T_c$ increases from $\sim$2 to 4-5 with increasing K doping level from optimal $x \sim$0.4 to strongly overdoped $x$=1.

\end{abstract}



\maketitle



\section{Introduction}

Superconductivity in (Ba$_{1-x}$K$_x$)Fe$_2$As$_2$ (BaK122 in the following) with transition temperature $T_c$ as high as 38~K was found by Rotter {\it et al.} \cite{Rotter} very soon after discovery of high-temperature superconductivity in LaFeAs(O,F) by Hosono group \cite{Hosono}. 
It was found later that superconductivity in BaFe$_2$As$_2$ can be also induced by electron doping on partial substitution of Fe atoms with aliovalent Co \cite{AthenaCo} and Ni \cite{Nidoping}, by isovalent substitution of Ru atoms at Fe sites \cite{Rudoping} and P atoms at As sites  \cite{Pdoping}, or by application of pressure \cite{pressure}.

In both families of compounds the superconductivity has maximum $T_c$ close to a point where the antiferromagnetic order of the parent compounds BaFe$_2$As$_2$ and LaFeAsO, respectively, is suppressed, prompting intense discussion about the relation of superconductivity and magnetism and potentially magnetic mechanism of superconducting pairing \cite{Paglione,CB,Johnston,Stewart,Louisreview}. 
A characteristic feature of the scenario, suggested for magnetically mediated superconductivity \cite{Lonzarich,Norman,Si}, is systematic doping evolution of all electronic properties, in particular of electrical resistivity. Superconducting $T_c$ has maximum at a point where line of the second order magnetic transition goes to $T=0$ (quantum critical point, QCP). Temperature dependent resistivity gradually transforms from $T^2$ expected in Fermi liquid theory of a metal away from QCP to $T$-linear at the QCP. In the transformation range $\rho(T)$ can be described with a second order polynomial, with the magnitude of $T$-linear scaling with superconducting $T_c$ \cite{Louisreview}. 
In iron-based superconductors this scenario works very well in iso-electron doped BaP122 \cite{Kasahara,YM,PQCP}. Here maximum $T_c$ is indeed observed at $x$=0.33, close to doping-tuned magnetic QCP, and signatures of QCP are found in both normal \cite{Kasahara,YM,PQCP} and superconducting \cite{HashimotoScience}  states, with resistivity at optimal doping being $T$-linear for both in-plane \cite{Kasahara} and inter-plane \cite{BaPcaxis} transport. Deviations from this scenario are not very pronounced in electron-doped BaCo122. Here maximum $T_c$ is observed close to a composition where $T_{N}(x)$ extrapolates to zero, though the actual line shows slope sign change on approaching $T$=0 and reentrance of the tetragonal phase \cite{Nandi}. The temperature-dependent in-plane resistivity is close to $T$-linear at optimal doping and transforms to $T^2$ in the overdoped regime, while the inter-plane resistivity shows limited range of $T$-linear dependence, terminated at high 
 temperature by a broad crossover \cite{anisotropy,anisotropypure,pseudogap,pseudogap2} due to pseudogap. The resistivity anisotropy $\gamma_{\rho} \equiv \rho_c/\rho_a$ scales with the anisotropy of the upper critical field $\gamma_H \equiv H_{c2,ab}/H_{c2,c}$ \cite{anisotropy} with $\gamma_{\rho}=\gamma_H^2$. The $\gamma_H(x)$ changes step-like between underdoped and overdoped regions of the dome \cite{AltrawnehCo,MurphyCo}, due to Fermi surface topology change (Lifshits transition) \cite{CoLifshits}.

Contrary to the cases of iso-electron substitution and electron doping, no systematic studies of the temperature-dependent resistivity and anisotropic properties of hole-doped BaK122 system were reported so far. Studies were performed in the underdoped, $x<0.4$ \cite{Wenresistivity,BaKcaxis} compositions, for which high quality single crystals can be grown from FeAs flux \cite{Wencrystals}, or in heavily overdoped range $x>0.76$ \cite{Watanabe}, where crystals were prepared from KAs flux \cite{Kihou}. Crystals of BaK122 can be also grown from Sn flux \cite{NiNiK}, however, their properties are notably affected by Sn inclusions at sub-percent level and will not be discussed here.

In BaK122 the superconductivity appears on sufficient suppression of antiferromagnetic order, for $x>\approx 0.15$, while magnetism is completely suppressed by $x \approx$0.25 \cite{Avci1,Avci2}, revealing a range of bulk coexistence. The doping edge of magnetism corresponds to $T_c\sim$27~K \cite{BaKcaxis}, notably lower that the highest $T_c \approx$38~K observed at optimal doping $x \approx$0.4, away from concentration boundary of magnetism suppression. The $T_c(x)$ dependence for $x$ in the range 0.4 to 0.6 is nearly flat \cite{21RotterAngCh}. The superconductivity is observed in the whole substitution range up to $x$=1 with steady decrease of $T_c$ down to 3.7~K in the end member KFe$_2$As$_2$ ($x$=1).

Broad crossover in the temperature dependent resistivity is observed in in-plane transport in single crystals of BaK122 at doping close to optimal \cite{Zverev}, similar to pure stoichiometric KFe$_2$As$_2$ (K122) \cite{Shiyan,Terashima,HashimotoKpure,ReidK,YLiu}. Explanation of the crossover was suggested as arising from multi-band effects \cite{Zverev}, with contribution of two conductivity channels, as found in optical studies \cite{Holms} with nearly temperature-independent and strongly temperature dependent resistivities, respectively. The maximum in $\rho_a(T)$ of BaK122 was discussed by Gasparov {\it et al.} \cite{Gasparov} as arising from phonon-assisted scattering between two Fermi-surface sheets.

The information about the doping-evolution of the upper critical field in hole-doped BaK122 is scattered. Very high upper critical fields were reported for close to optimally doped compositions \cite{Yuan,WenHc2,Gasparov}, in addition these compositions are characterized by rather small critical field anisotropy. In another doping regime, close to $x$=1, very unusual behavior of the upper critical fields is found. In KFe$_2$As$_2$, the orbital $H_{c2}$ found in $H \parallel c$ configuration, is close to $T$-linear \cite{Terashima}. The slope of the dependence does not depend on $T_c$ suppression with impurities \cite{YLiu}. In configuration with magnetic field parallel to the plane, $H\parallel ab$, the upper critical field is Pauli limited, as suggested both by the difference in the shape of the phase diagram and quite sharp changes at $H_{c2}$ \cite{Lohneissen}. Heat capacity study in $H \parallel a$ configuration, however, had not found first order transition\cite{Kittaka}, but rather suggested multi-band Fulde-Ferrel-Larkin-Ovchinnikov (FFLO \cite{LO,FF}) state \cite{Machida}. In slightly less doped material with $x$=0.93, ($T_c \sim$8 K) hysteresis is observed in the field-tuned resistive transition curves in $H \parallel ab$ configuration at temperatures below 1 K, which can be attributed to a first-order superconducting transition due to paramagnetic effect \cite{TerashimaBaKHc2}. More systematic studies of the anisotropic $H_{c2}$ in BaK122 system are desperately required.

In this study, we report growth of high quality single crystals of (Ba$_{1-x}$K$_x$)Fe$_2$As$_2$ for all doping ranges ($0.22 \leq x \leq 1$) and report systematic study of their temperature-dependent resistivity and anisotropic upper critical fields.  We found nearly doping independent resistivity value at high temperatures, which is in notable contrast to electron-doped BaCo122 \cite{pseudogap} and iso-electron substituted BaP122 \cite{Kasahara} materials. We find systematic evolution of the temperature dependent resistivity with doping and rapid decrease of residual resistivity towards $x$=1. We also found that the slopes $ dH_{c2}/dT$ are proportional to $T_c$ for both $H \parallel c$ and $H \parallel ab$ configurations. The anisotropy $\gamma \equiv H_{c2,ab}/H_{c2,c}$, increases from ~2 to 4-5 with increasing K doping level. The doping dependence of anisotropy ratio might be linked with change of the topology of the Fermi surface and the evolution of the superconducting gap.

\section{Experimental}

\subsection{Crystal growth}

\begin{figure}[tbh]%
\centering
\includegraphics[width=7cm]{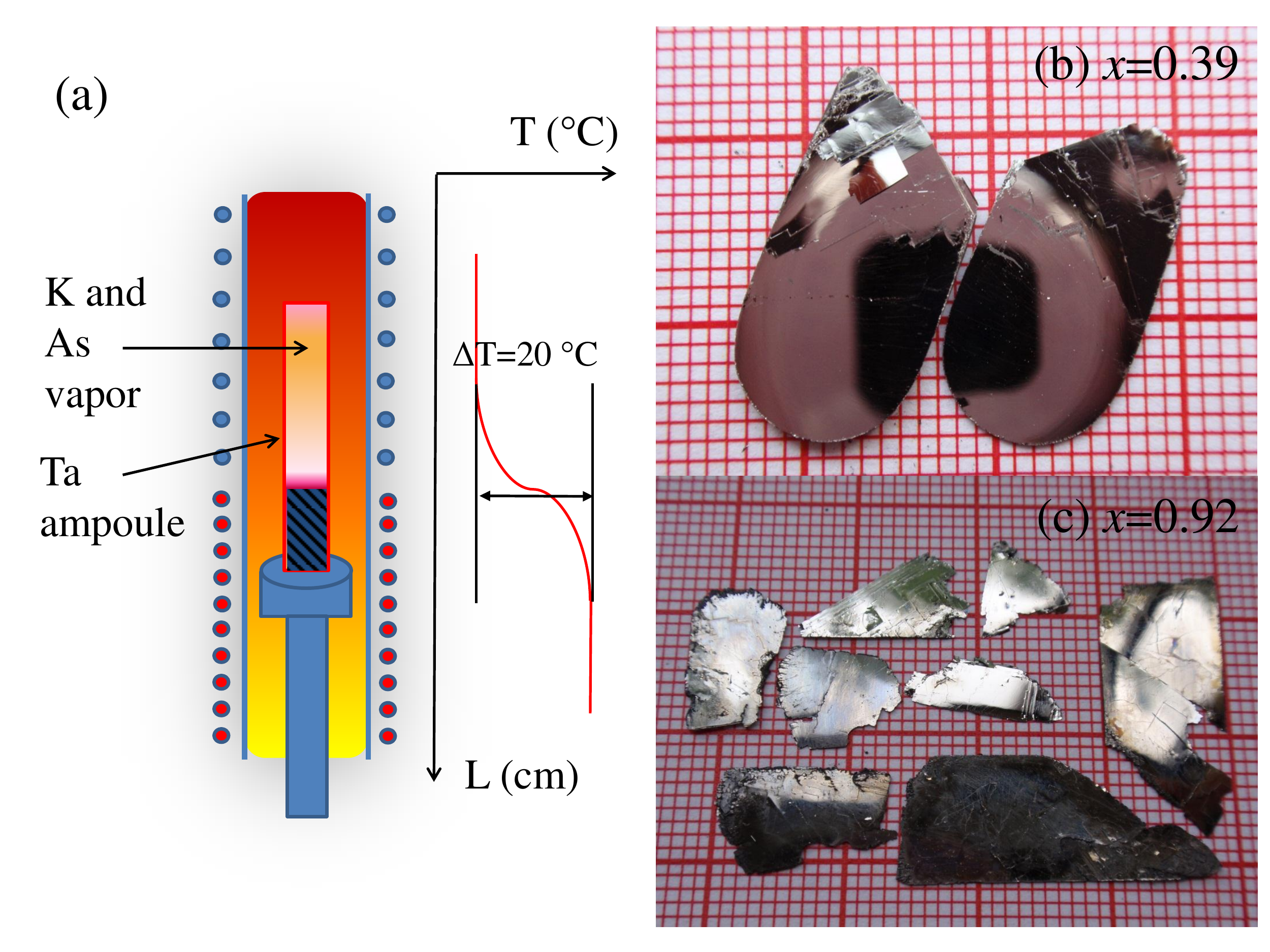}%
\caption{ (a) Single crystals of (Ba$_{1-x}$K$_x$)Fe$_2$As$_2$ were grown in a vertical two-zone tube furnace, in which the temperature of the top zone during the growth was set 20K lower than the temperature of the bottom zone. (b) Photograph of the as-grown single crystal with $x$=0.39 cleaved along the $ab$ plane, showing sample with in-plane dimensions up to 18$\times$10 mm$^2$. The reflection of the camera can be seen in the mirror-like surface. (c) Thin plate-like crystal of heavily K doped composition $x$=0.92 with size up to 15$\times$10 mm$^2$.
}%
\label{growth}%
\end{figure}

We have previously described successful growth of the high quality single crystals of stoichiometric KFe$_2$As$_2$ at $T$=1157~K using KAs flux \cite{YLiu}. One of the key elements of the growth technique was using a liquid-Sn sealing of alumina crucibles to suppress the evaporation of K and As. This technique allowed us to avoid use of quartz tubes in direct contact with K vapor and use of expensive sealed tantalum tubes. Analysis of the growth morphology in the case of KFe$_2$As$_2$ \cite{YLiu} lead us to conclude that the crystals nucleate on the surface of the melt and grow by the reaction on the top surface of the crystal with K and As in the vapor phase. We were able to promote this reaction by developing an inverted-temperature-gradient method with the colder zone at the top of the crucible, as shown in Fig.~\ref{growth}(a). A temperature gap of 20 K was set between the top zone and the bottom zone. This method yielded higher quality crystals of KFe$_2$As$_2$ with residual resistivity ratio of up to 3000 than obtained in traditional flux-method, as crystallization from the liquid top can expel impurity phases into the liquid during crystal growth.

This method works very well for the growth of heavily K doped (Ba$_{1-x}$K$_x$)Fe$_2$As$_2$ single crystals. Small amount of Ba was added to the load with the ratio Ba:K:Fe:As=$y$:5:2:6 ($y $=0.1, 0.2, and 0.3) in the stating materials. The chemicals were weighed and loaded into an alumina crucible in a glove box under argon atmosphere. Because of use of higher soaking temperatures leading to higher vapor pressures Sn seal technique was not reliable enough, and we switched to tantalum tube sealing.  The alumina crucibles were then sealed in a tantalum tube by arc welding. In Table~\ref{table1} we show the growth conditions of Ba$_{1-x}$K$_x$Fe$_2$As$_2$ single crystals. We started to grow heavily K doped crystals by following the same procedure that had worked well for the crystal growth of KFe$_2$As$_2$. For $y$ =0.1, we obtained single crystals with K doping level at around $x$=0.90 using soaking temperature of 1193 K. The actual compositions of the crystals were determined by wavelength dispersive x-ray spectroscopy (WDS) electron-probe microanalysis. For $y$ =0.2 and 0.3, the single crystals obtained by cooling down from the soaking temperature of $T$=1193 K display broad transitions, which suggests inhomogeneity of Ba and K distributions in the sample. We were able to improve sample quality by adjusting the composition of the starting load material and soaking temperatures, as shown in Table~\ref{table1}. We found that increase of the soaking temperature to $T$=1273~K helps growth of the samples with $x$=0.8 and 0.9 with sharp superconducting transition. The further increase of the soaking temperature up to 1323~K, leads to growth of the crystals showing multiple steps at the superconducting transition due to inhomogeneous K distribution. We found that higher soaking temperatures 1273K$\leq T \leq$1323 K and narrowed temperature window for crystal growth are similarly useful to grow the crystals within the doping range $0.6<x<0.9$ with sharp superconducting transition.

\begin{table*}
\caption{\label{table1}Growth conditions of (Ba$_{1-x}$K$_x$)Fe$_2$As$_2$ single crystals. Soaking temperature corresponds to the set temperature of bottom zone, with the top zone temperature 20 K lower than the bottom zone.}
\vspace{.2cm}
\begin{tabular}{lccccc}
\hline
Target K doping level  & Starting mixture   & Soaking temperature & Soaking time & Cooling rate      	\\ 
\hline
$x$=0.22        & Ba:K:Fe:As= 1-$x$:$x$:6:6&	 1453 K& 2h & 2 K/h to 1313 K 		\\
\hline
\vspace{.1cm}

$x$=0.34, 0.39, 0.47, 0.53 & Ba:K:Fe:As= 1-$x$:2$x$:4:5&	 1413 K & 2h & 1 K/h to 1293 K \\

\hline
\vspace{.1cm}

$x$= 0.55 & Ba:K:Fe:As= $x$:3$x$:4:5&	 1393 K & 2h & 0.5 K/h to 1293 K \\

\hline
\vspace{.1cm}

$x$= 0.65, 0.80 and 0.82 &  Ba:K:Fe:As=$y$:4:2:5&	 1273  K & 6h & 4 K/h to 1173 K\\
&(Ba: y=0.2~0.3) & & & 1 K/h to 973 K\\

\hline
\vspace{.1cm}

$x$=0.90 and 0.92& Ba:K:Fe:As=$y$:4:2:5 & 1273  K & 2h & 3 K/h to 1173 K
\\
                     & Ba: $y$=0.1  &         &   & 1 K/h to 973 K\\
 \hline
\vspace{.1cm}
$x$=1 & K:Fe:As=5:2:6 & 1193 K & 1h & 4 K/h to 1093 K\\

\hline
\vspace{.1cm}

\label{samples}
\end{tabular}
\end{table*}



For the samples with K doping levels below $x$=0.55, we turned to the FeAs flux method. The growth conditions can be found in Table~\ref{table1}. For the crystals within the optimal doping range ($0.3<x<0.5$), the growth using conditions as shown in Table~\ref{table1} yielded large and high quality crystals with sharp transition. Interestingly, to grow high quality underdoped crystals, a further increase of the soaking temperature to 1453 K and fast cooling rate of 2 K/h are needed. A series of large and high quality (Ba$_{1-x}$K$_x$)Fe$_2$As$_2$  single crystals ($0.22 \leq x \leq1 $) with sizes up to $18 \times 10\times 1$ mm$^3$, as shown in Fig.~\ref{growth}(b) for $x$=0.39 and Fig.~\ref{growth} (c) for $x$=0.92. In fact, the size of Ba$_{1-x}$K$_x$Fe$_2$As$_2$ single crystals was only limited by the size of alumina crucibles used.

\subsection{ Sample characterization}

\begin{figure}[tbh]%
\centering
\includegraphics[width=7cm]{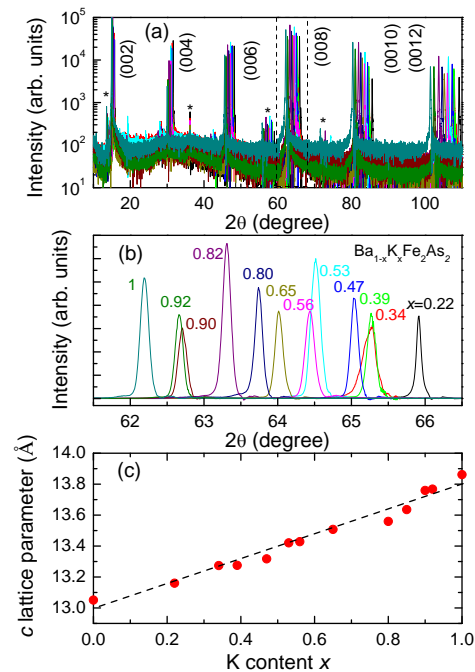}%
\caption{  X-ray diffraction patterns of (Ba$_{1-x}$K$_x$)Fe$_2$As$_2$ $0.22 \leq x \leq 1$ single crystals. The logarithmic plot reveals small amount of impurity phases indicated by the asterisks, which could result from the FeAs and KAs flux inclusions. (b) The (008) peak, seen in 61$^{\circ} <2 \Theta <67 ^{\circ}$ range, systematically shifts with increasing K doping level towards the low angles. (c) The $c$ lattice parameter changes linearly the K content $x$. The dashed line is guide for eyes.
}%
\label{Xray}%
\end{figure}

XRD measurements were performed on a PANalytical MPD diffractometer using Co $K \alpha $ radiation. The $K \alpha 2$ radiation was removed with X'pert Highscore software. All BaK122 crystals are readily cleaved along the $ab$ plane, as shown in Figs. ~\ref{growth}(b)-(c). The XRD patterns of BaK122 single crystals with $0.22 \leq x \leq 1$ are shown in Fig.~\ref{Xray}. The traces of impurity phases close to the baseline are indicated by the asterisks, they are most likely caused by the flux inclusions.  Figure~\ref{Xray}(b) shows systematic shift of the (008) peak towards the lower angles with increasing K content. The $c$-axis lattice parameter is estimated based on the (00$l$) diffractions and displayed as a function of K content in Fig.~\ref{Xray}(c); it changes linearly with $x$ its values match well the results on polycrystalline samples \cite{21RotterAngCh}.

Magnetic susceptibility $\chi(T)$ was measured using PPMS Vibrating Sample Magnetometer ({\it PPMS VSM, Quantum Design}). Typical size of the single crystals used in magnetization measurements was 4$\times$3$\times$0.2 mm$^3$, and their mass was $\sim~10$ mg. In-plane resistivity $\rho_{a}$ was measured in four-probe configuration using Physical Property Measurement System ({\it PPMS, Quantum Design}). Samples were cleaved into bars with typical dimensions (1-2)$\times$0(0.3-0.5)$\times$(0.02-0.05) mm$^3$. Electrical contacts were made by soldering Ag wires using pure tin \cite{SUST,patent} and had contact resistance typically in several $\mu \Omega$ range. Sample dimensions were measured using optical microscope with the accuracy of abou 10\%. Quantitative characterization of resistivity was made on a big array of samples of each composition.

\begin{figure}[tbh]%
\centering
\includegraphics[width=7cm]{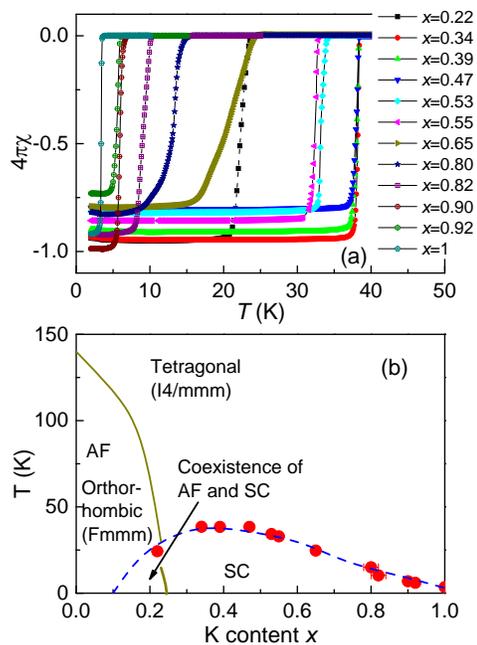}%
\caption{  (Color online) (a) Temperature dependence of the magnetic susceptibility $\chi(T)$ of (Ba$_{1-x}$K$_x$)Fe$_2$As$_2$ $0.22 \leq x \leq 1$ single crystals. Bulk superconducting transition temperature $T_c$ was determined from the onset point of the rapid drop of $\chi (T)$. (b) Doping phase diagram of (Ba$_{1-x}$K$_x$)Fe$_2$As$_2$ as determined from magnetization measurements on single crystals $0.22 \leq x \leq 1$. The superconducting transition temperature (red solid dots), $T_c(x)$, matches well that obtained on polycrystalline samples (blue dashes) \cite{21RotterAngCh,Avci1,Avci2}. Solid line shows boundary of orthorhombic/antiferromagnetic phase from neutron scattering study on polycrystals \cite{21RotterAngCh,Avci1,Avci2}. }%
\label{phaseD}%
\end{figure}

In Fig.~\ref{phaseD} we show the temperature dependence of magnetic susceptibility $\chi(T)$ for BaK122 single crystals with $0.22 \leq x \leq 1$. Sharp superconducting transition ($\Delta T_c<$0.6~K) in magnetic susceptibility curves show high quality of crystals with $x$=0.34, 0.39, 0.47, 0.53, 0.55, and 1. The transition width $\Delta T_c$ was defined using 90\% and 10\% drop in $\chi(T)$ of the full diamagnetic transition as the criterion. The samples with $x$=0.82, 0.90, and 0.92 have $\Delta T_c<$1 K. However, the samples with $x$=0.65 and 0.80 have large $\Delta T_c$ of 3 K and 5 K, respectively. As we mentioned in the Experimental section, we shifted the temperature windows and adjusted the starting load composition and materials to improve the sample quality and obtain sharper transitions.

Using $T_c$ from magnetic susceptibility data of top panel of Fig.~\ref{phaseD} and $x$ values as obtained in WDS analysis, we constructed the doping phase diagram, as shown in bottom panel of Fig.~\ref{phaseD}. For reference we show the diagram as determined from measurements on high quality polycrystalline materials \cite{Avci2} and on high quality single crystals on the underdoped side \cite{Wencrystals}. The three studies are in good agreement.

We do not see any indications of the phase separation in our underdoped samples $x$=0.22. Previous study of underdoped BaK122 samples grown from Sn flux with $x$=0.28 found regions of antiferromagnetically (AF) ordered phase with size of 65 nm coexisting with nonmagnetic superconducting regions \cite{19Park}. Later study using three-dimensional (3D) atom probe tomography revealed that the separation is caused by inhomogeneous distributions of Ba and K elements \cite{20Yeoh}, with a tendency for Ba and K atoms to form clusters. Thus we conclude that this problem is not characteristic of the growth technique we use.

We do see, however, that strong inhomogeneity occurs during crystal growth of overdoped crystals ($0.65<x<0.8$). There is no intrinsic phase separation revealed for polycrystalline samples in this doping range. In our samples we do not see macroscopic inhomogeneity in WDS measurements with special resolution of about 1 $\mu m$. However, on finer scale two STM studies revealed ordered vortex lattice in single crystals of optimally doped BaK122 $x$=0.40 \cite{LShan} but a short-range hexagonal order (vortex glass phase) in single crystals of SrK122 $x=$0.25 \cite{Song}. Song {\it et al.} suggested that mismatch between the size of the dopant K atom and of the host atoms Ba and Sr, Ba$^{2+}$/ K$^+$ and Sr$^{2+}$/ K$^+$, respectively, causes dopant clustering, electronic inhomogeneity, and vortex glass phase in SrK122. K$^+$ ions should be less clustered in BaK122 than in SrK122 because ion size mismatch between K$^+$ and Ba$^{2+}$ is five times smaller than between K$^+$ and Ba$^{2+}$  \cite{Song}. Thus one can expect that dopant (Ba$^{2+}$/ K$^+$) size mismatch diminishes for $x$=0.50 doped sample, which is quite close to the optimally doped sample with $x$=0.40. Detailed structure analysis \cite{Avci2} indicates that potassium substitution reduces the in-plane lattice parameters $a$ and $b$ and significantly increases the out-of-plane lattice parameter $c$ in BaK122 compounds, all measured at $T$=1.7 K. The unit cell volume gradually decreases with increasing K content until $x\approx$ 0.5, but slightly increases towards the end member KFe$_2$As$_2$. Although the evolution of the lattice parameters and of the unit cell volume does not provide us direct evidence about the lattice mismatch in the BaK122 compounds, the slight change of the unit cell volume for doping in the range 0. 5$\leq x \leq$1 implies that Ba$^{2+}$/K$^+$ ion clustering should be relatively easy to realize without too much disturbance in crystal structure in the overdoped samples. In other words, a broad distribution of K doping in the overdoped samples should not be challenged too much when considering weak lattice strains introduced by different K dopings. Therefore, it is important to search for suitable growth conditions to narrow K doping range when growing overdoped single crystals.

\section{Results and discussion}

\subsection{ Doping evolution of the temperature-dependent resistivity }

\begin{figure}[tbh]%
\centering
\includegraphics[width=7cm]{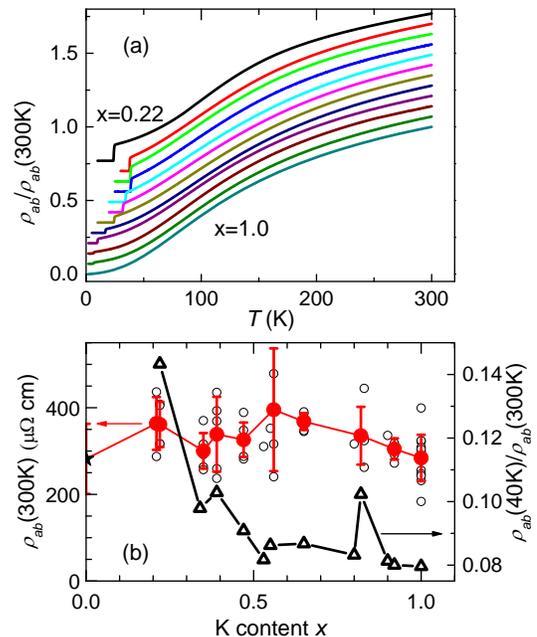}%
\caption{  (Color online) (a) Temperature dependence of in-plane resistivity in single crystals of (Ba$_{1-x}$K$_x$)Fe$_2$As$_2$ (top to bottom $x=$0.22, 0.34, 0.39, 0.47, 0.53, 0.55, 0.65, 0.80, 0.82, 0.90, 0.92, 1.0). The data are presented using normalized $\rho (T)/\rho (300K)$ plot and offset for clarity;  (b) (right scale) Doping evolution of the $\rho(40K)/\rho(300K)$ and (left scale) of the room-temperature resistivity $\rho (300K)$, open dots are data for individual  samples, solid dots with error bars show statistical average and standard deviation. }%
\label{resistivity}%
\end{figure}

\begin{figure}[tbh]%
\centering
\includegraphics[width=7cm]{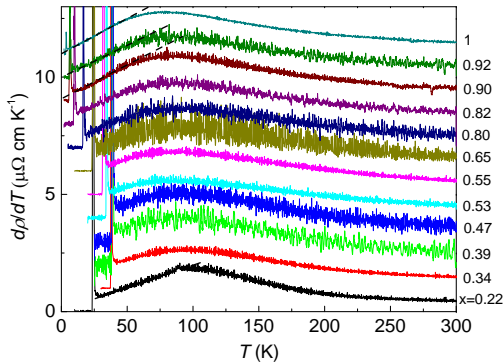}%
\caption{  (Color online) Doping evolution of the temperature- dependent derivative of in-plane resistivity, 
$d\rho_a(T)/dT$,  in single crystals of (Ba$_{1-x}$K$_x$)Fe$_2$As$_2$ $0.22 \leq x \leq 1$. The data are offset to avoid overlapping. }%
\label{derivative}%
\end{figure}

Temperature dependent in-plane resistivity $\rho(T)$ of the samples with $x$= 0.22 to 1.0 is shown in Fig.~\ref{resistivity}. The data are presented using normalized $\rho(T)/\rho(300K)$ plots and offset to avoid overlapping. The doping evolution of the actual resistivity values $\rho(300K)$ shows significant scatter due to uncertainty of the geometric factors, which are strongly affected by hidden cracks in micacious crystals of iron pnictides \cite{AltrawnehCo,anisotropy}. Of note though that within statistical error, the resistivity $\rho(300K)$ remains constant over the whole compositional range from heavily underdoped samples with $x$=0.22 to heavily overdoped $x$=1.0, which is distinctly different from electron doped BaCo122 \cite{Alloul,pseudogap} and isoelectron substituted BaP122 \cite{Kasahara}, in which $\rho(300K)$ decreases notably with doping. The first look at the temperature-dependent resistivity also does not show significant doping evolution. For all doping 
 levels the $\rho(T)$ curves show a broad crossover starting above 100~K and ending at around 200~K. The onset of this feature most clearly reveals itself as a maximum in the temperature-dependent resistivity derivative, see Fig.~\ref{derivative}. The origin of the feature was discussed in terms of multi-band character of conductivity in which one of the bands has strongly temperature dependent contribution, while the other has nearly temperature independent conductivity \cite{Zverev}, as contribution from phonon-assisted scattering between two Fermi-surface sheets \cite{Gasparov} and as a feature associated with pseudogap, as suggested by its correlation with the maximum of the inter-plane transport $\rho_c (T)$ in under-doped compositions \cite{BaKcaxis,pseudogap,pseudogap2}. The position of the crossover does not change with doping, and since the Fermi surface topology reveals quite significant changes \cite{ARPESFS}, the explanation of the maximum in term of special featu
 res of band structure \cite{Zverev,Gasparov} is very unlikely.

\begin{figure}[tbh]%
\centering
\includegraphics[width=7cm]{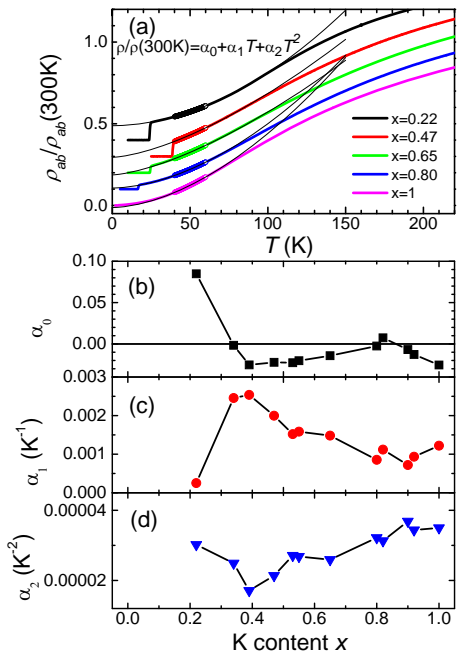}%
\caption{  (Color online) (a) Top panel- fixed 40 to 60K range fit of the resistivity curves using second order polynomial $\rho/\rho(300K)=\alpha_0+\alpha_1*T+\alpha_2*T^2$, shown for selected dopings $x$=0.22, 0.47, 0.65, 0.8, 1. The data are offset to avoid overlapping. Three panels at the bottom show doping evolution of the fit parameters $\alpha_0$ (panel b), $\alpha_1$ (panel c) and $\alpha_2$ (panel d). }%
\label{fit}%
\end{figure}

\begin{figure}[tbh]%
\centering
\includegraphics[width=7cm]{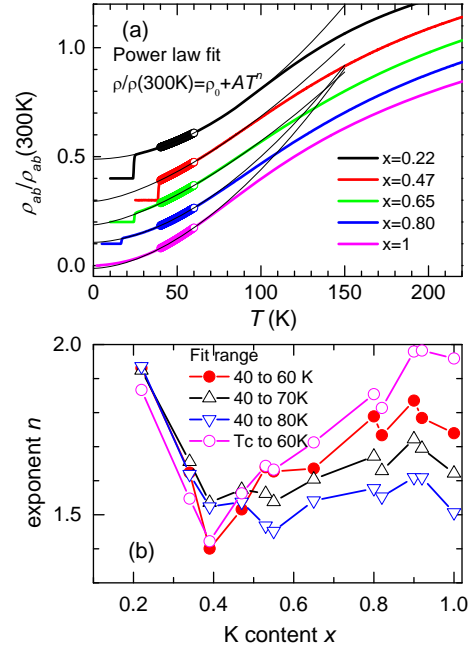}%
\caption{  (Color online) Top panel- fixed 40 to 60K range fit of the resistivity curves using power-law function $\rho/\rho(300K)=\rho_0+AT^n$, shown for selected dopings $x$=0.22, 0.47, 0.65, 0.8, 1. The data are offset to avoid overlapping. Bottom panel shows evolution of the power-law exponent $n$ with doping for fits over four different temperature ranges, 40 to 60 K as shown in top panel (red solid circles), 40 to 70~K (black up-triangles), 40 to 80~K ((blue down-triangles) and $T_c$ to 60~K (magenta open circles). }%
\label{fit2}%
\end{figure}

\begin{figure}[tbh]%
\centering
\includegraphics[width=7cm]{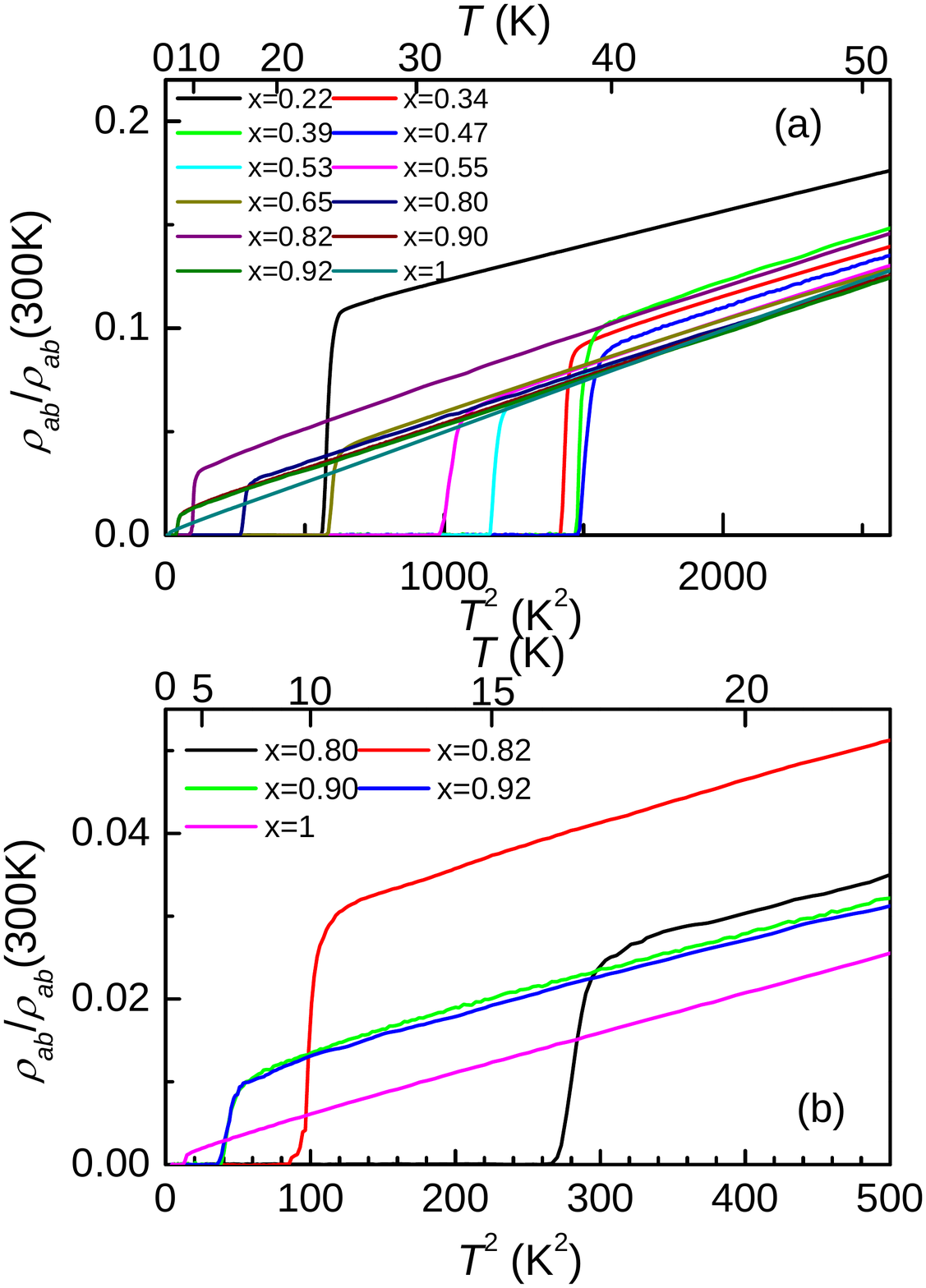}%
\caption{  (Color online) Top panel- normalized resistivity curves $\rho/\rho(300K)$ plotted vs $T^2$ for all doping levels studied, $x=$0.22, 0.34, 0.39, 0.47, 0.53, 0.55, 0.65, 0.80, 0.82, 0.90, 0.92, 1.0. Bottom panel shows data over narrower temperature range in heavily overdoped compositions $x$=0.80, 0.82, 0.90, 0.92, 1.0. The data does not show any significant dependence of the slope (proportional to $T^2$ coefficient $\alpha_2$ even at low temperatures. 
}%
\label{Tsquare}%
\end{figure}

At temperatures lower than 100~K, however, temperature-dependent resistivity shows some evolution. Because of high temperature of the superconducting transition, we cannot make correct analysis of the functional form of $\rho(T)$ in the $T \to 0$ limit over the whole dome. However, for the sake of comparison, we fitted the curves in a narrow range from 40 to 60~K, which was fixed for all compositions. These fits were done two ways. The first approach was using second order polynomial function, $\rho(T)/\rho(300K)=\alpha_0 +\alpha_1*T+ \alpha_2*T^2$, similar to the fit used by Doiron-Leyraud {\it et al.} \cite{NDL} for electron-doped BaCo122. In the top panel of Fig.~\ref{fit} we show the fits over the range 40 to 60 K for $\rho(T)$ curves for representative doping levels, three bottom panels show doping evolution of the fit parameters $\alpha_0$, $\alpha_1$ and $\alpha_2$. This analysis reveals clearly that the dependence has highest linear contribution at $x$=0.35 and 0.39, and that the $T^2$ contribution is minimum at $x$=0.39, coinciding with maximum $T_c$ position but away from the doping border of the antiferromagnetic state at $x$=0.26.

The second approach was fitting the data using a power-law function, $\rho/\rho(300K)=\rho_0+AT^n$, as shown for selected compositions in Fig.~\ref{fit2}. This approach is similar to the approach used by Shen {\it et al.} \cite{Wenresistivity}, however in their case the fitting range was extending to 80~K. For the sake of comparison, we did power-law analysis for the temperature ranges 40 to 70~K and 40 to 80~K, and from above $T_c$ to 60~K. The results of these fittings are shown in the bottom panel of Fig.~\ref{fit2}. It can be seen that all ways of analysis find largest deviations from Fermi-liquid $T^2$ dependence at $x$=0.39, which corresponds to a leading edge of maximum $T_c$ plateau of the $T_c(x)$ dome. Since $T_c(x)$ function is flat in 0.34 to 0.56 range, while both $T$-linear contribution in the polynomial analysis, Fig.~\ref{fit}(c), and power-law exponent $n$ peak at $x$=0.39, we conclude that $T_c$ and the amplitude of $T$-linear contribution do not scale 
 in BaK122, contrary to BaCo122 \cite{Louisreview}. 
Another interesting point is that exponent $n$ we observe in sample $x$=0.39 is close to 1.5. This is notably higher than the lowest exponent $n$=1.1 found in previous study \cite{Wenresistivity}.
To further check the link between $T$-linear contribution and maximum $T_c$, further studies in high magnetic fields may be necessary. 

An interesting feature of these fits is that the residual resistivity takes negative values for most of the compositions. This fact is suggestive that at lower temperatures the $\rho(T)$ curves should develop significant positive curvature, as is in fact observed for heavier doped compositions, in which broader temperature range can be studied. It also suggests that most of our samples have quite high residual resistivity ratio in $T \to 0$ limit.

On the other hand, the $T^2$ coefficient as determined from the polynomial fit for the range 40 to 60~K gradually increases towards $x$=1. Since $T_c$ drops significantly in this range, we are able to make an analysis at lower temperatures. In Fig.~\ref{Tsquare} we plot $\rho (T)$ data for all samples using a $T^2$ plot, bottom panel shows expanded view for heavily overdoped samples. When plotted this way, the plots become linear right above $T_c$, and the slopes of the curves do not show any noticeable doping evolution beyond error bars. This observation suggests that for all doping levels there is significant and non-critical $T^2$ coefficient, and indeed several contributions to conductivity are needed for correct account of its doping evolution.

\subsection{ Anisotropic upper critical fields} 

The anisotropy of the upper critical field $\gamma_H \equiv \frac{H_{c2ab}}{H_{c2c}}$ presents important information about the anisotropy of the electrical conductivity, $\gamma_{\rho} \equiv \frac{\rho_c}{\rho_a}$. In a temperature range close to zero-field $T_c$ the two anisotropies are related as $\gamma_H^2 = \gamma_{\rho}$, a relation which was verified semi-quantitatively in KFe$_2$As$_2$  \cite{Terashima}. The angular dependent $H_{c2}(\Theta)$ was also studied systematically in BaK122 with $x$=0.92 \cite{TerashimaBaKHc2}, in which the authors found strong deviations from $\cos(\Theta)$ dependence expected in orbital limit \cite{MurphyNi}. Scattered in $x$ measurements of $\gamma_H$ were undertaken on samples close to optimal doping grown from Sn flux \cite{NiNiK,NiNiK2,Gasparov,Yuan} and from FeAs flux \cite{WelpK,WenHc2}. Here we study evolution of the $\gamma_H (x)$ in BaK122 from resistive $H_{c2}$ measurements. 

\begin{figure}[tbh]%
\centering
\includegraphics[width=7cm]{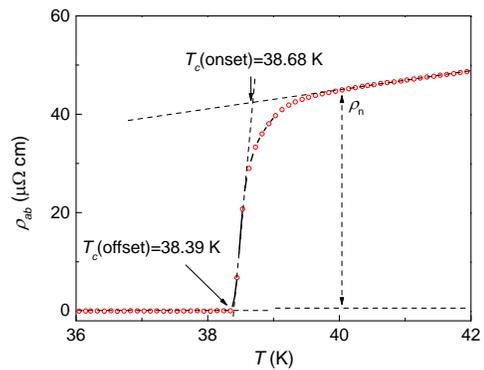}%
\caption{  (Color online) Temperature-dependent resistivity in single crystal of (Ba$_{1-x}$K$_x$)Fe$_2$As$_2$ with $x$=0.39 in the vicinity of the superconducting transition. The onset $T_{c,onset}$ of the transition is defined at the crossing point of the linear fits of the $\rho(T)$ in the normal state above $T_c$ and at the sharp transition slope. The offset $T_c$ corresponds to the extrapolation of the steep transition slope to zero resistance. }%
\label{Tcdefinitions}%
\end{figure}

In Fig.~\ref{Tcdefinitions} we show zoom of the $\rho (T)$ curve in the vicinity of the superconducting transition in sample with $x$=0.39. Here we show how we defined different criteria used to determine $T_c(H)$ dependence.  We analyzed resistivity data by linear extrapolation of $\rho (T)$ curves at the transition and above the transition. The onset $T_{c,onset}$ of the transition is defined at the crossing point of these linear fits.  The offset $T_c$ corresponds to the crossing of the steep transition line with $\rho =0$ line. 


\begin{figure*}[tbh]%
\centering
\includegraphics[width=10cm]{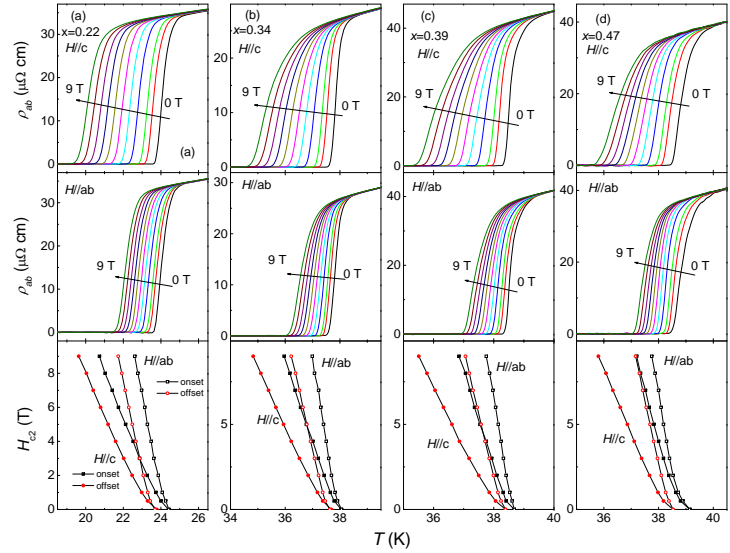}%
\caption{ (Color online)
Temperature dependence of in-plane resistivity in single crystals of (Ba$_{1-x}$K$_x$)Fe$_2$As$_2$ with $x$=0.22 (a), 0.34 (b), 0.39 (c) and 0.47 (d) in magnetic fields $H \parallel c$ (top panels) and $H \parallel ab$ (middle panel) with magnetic fields (right to left) 0, 0.5, 1, 2, 3, ..., 9 T. Bottom panels show $H_{c2}(T)$ for two field orientations $H \parallel c$ (solid symbols) and $H \parallel ab$ (open symbols) as determined using onset (black squares) and offset (red circles) resistive transition criteria, see Fig.~\ref{Tcdefinitions}.}%
\label{Hc2rowunder}%
\end{figure*}

\begin{figure*}[tbh]%
\centering
\includegraphics[width=10cm]{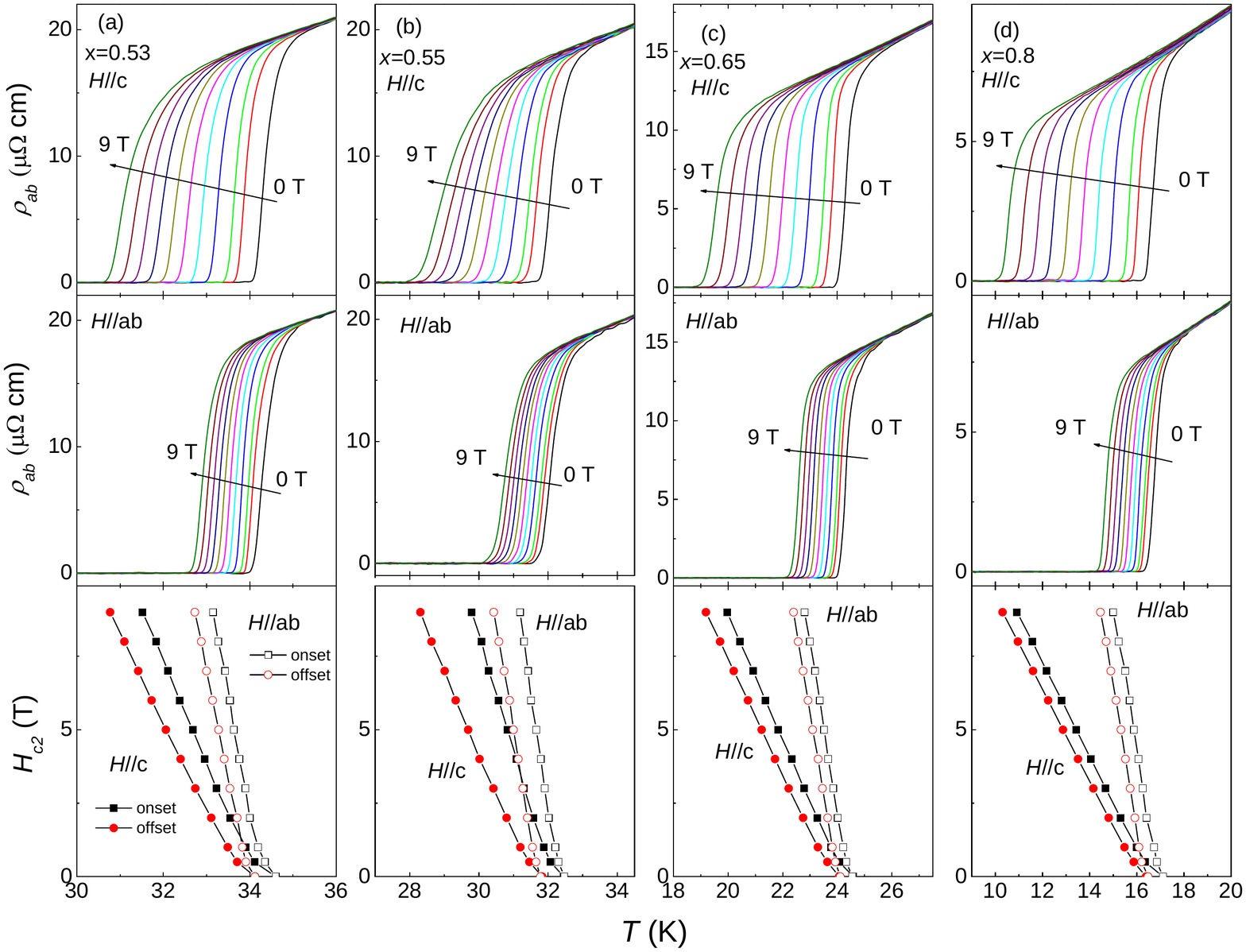}%
\caption{  (Color online) Temperature dependence of in-plane resistivity in single crystals of (Ba$_{1-x}$K$_x$)Fe$_2$As$_2$ with $x$=0.53 (a), 0.55 (b), 0.65 (c) and 0.80 (d) in magnetic fields $H \parallel c$ (top panels) and $H \parallel ab$ (middle panel) with magnetic fields (right to left) 0, 0.5, 1, 2, 3, ..., 9 T. Bottom panels show $H_{c2}(T)$ for two field orientations $H \parallel c$ (solid symbols) and $H \parallel ab$ (open symbols) as determined using onset (black squares) and offset (red circles) resistive transition criteria, see Fig.~\ref{Tcdefinitions}.}%
\label{Hc2rowslight}%
\end{figure*}

\begin{figure*}[tbh]%
\centering
\includegraphics[width=10cm]{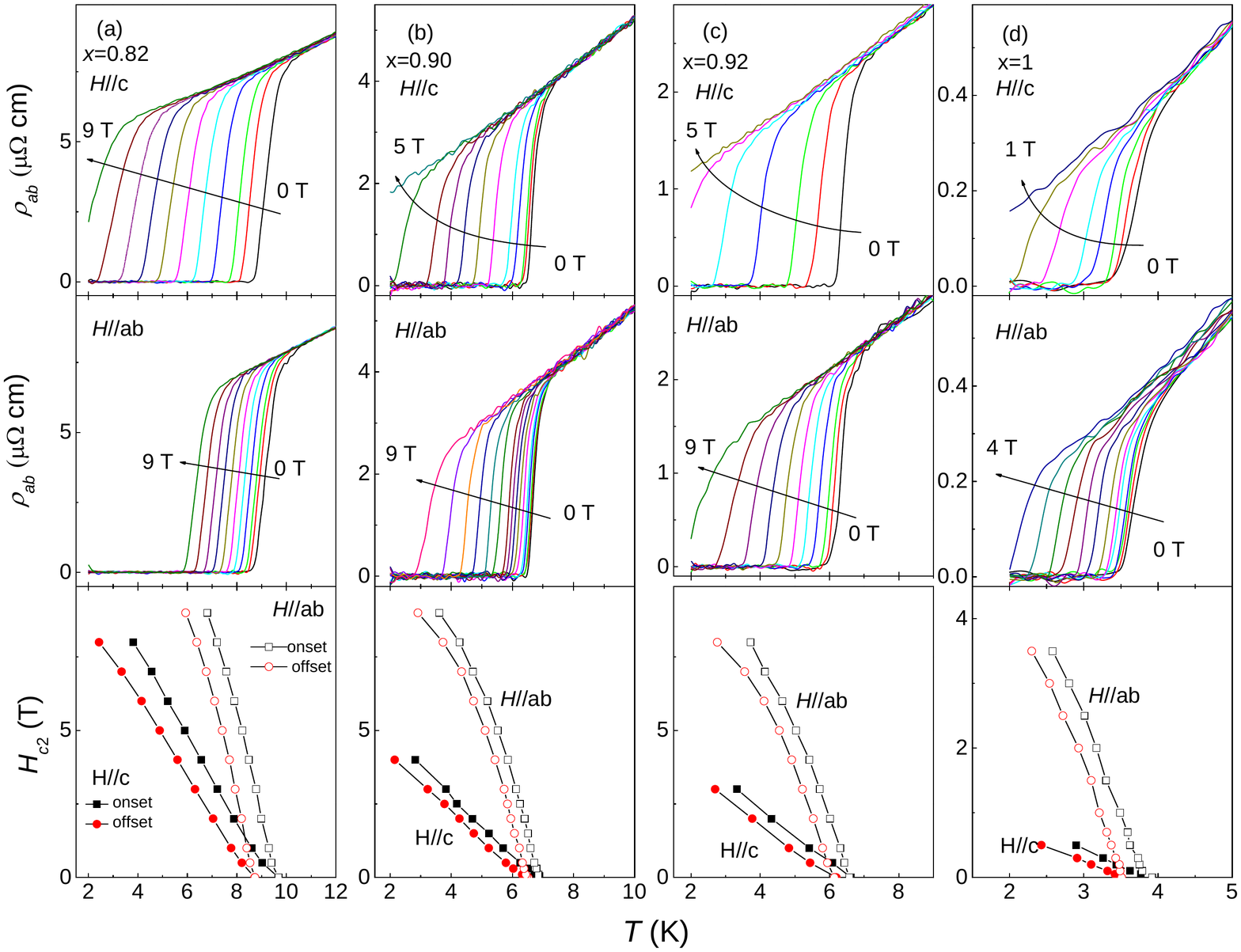}%
\caption{   (Color online) (Top panels) Temperature dependence of in-plane resistivity in single crystals of (Ba$_{1-x}$K$_x$)Fe$_2$As$_2$ with $x$=0.82 (a) (right to left 0, 0.5, 1, 2, 3, ..., 9 T), 0.90 (b) (right to left 0, 0.05, 0.1, 0.3, 0.5, 1, 1.5, 2, 2.5, 3, 4, 5 T), 0.92 (c) (right to left 0, 0.5, 1, 2,..., 5 T) and 1 (d) (0, 0.05, 0.1, 0.2, 0.3, 0.5, 0.7, 1 T) in magnetic fields $H \parallel c$. Middle panels show the data for $H \parallel ab$ (a), $x$=0.82, right to left 0, 0.5, 1, 2, 3, ..., 9 T, (b) $x$=0.90, field values right to left 0, 0.05, 0.1, 0.3, 0.5, 1, 1.5, 2, 2.5, 3, 4, ..., 9 T, (c) $x$=0.92 right to left 0, 0.5, 1, 2, 3, ..., 9 T and (d) $x$=1 magnetic fields 0, 0.1, 0.2, 0.3, 0.5, 0.7, 1, 1.5, 2, 2.5, 3, 3.5, 4 T. Bottom panels show $H_{c2}(T)$ for two field orientations as determined using onset (squares) and offset (circles) of of resistive transition criteria, see Fig.~\ref{Tcdefinitions}.}%
\label{Hc2rowheavy}%
\end{figure*}

In Fig.~\ref{Hc2rowunder} we show resistivity data taken in magnetic fields parallel to $c$-axis (top panels), parallel to the conducting $ab$ plane (middle panels) and temperature dependent $H_{c2}(T)$ for two field orientations determined using onset and offset criteria. The data are shown for BaK122 compositions with $x$=0.22 (a), 0.34 (b), 0.39 (c), and 0.47 (d). Similar data for slightly to moderately overdoped compositions $x$= 0.53 (a), 0.55 (b), 0.65 (c), 0.80 (d) are shown in Fig.~\ref{Hc2rowslight}, and for strongly overdoped compositions $x$=0.82 (a), 0.90 (b), 0.92 (c) and 1.0 (d) in Fig.~\ref{Hc2rowheavy}.


\begin{figure}[tbh]%
\centering
\includegraphics[width=7cm]{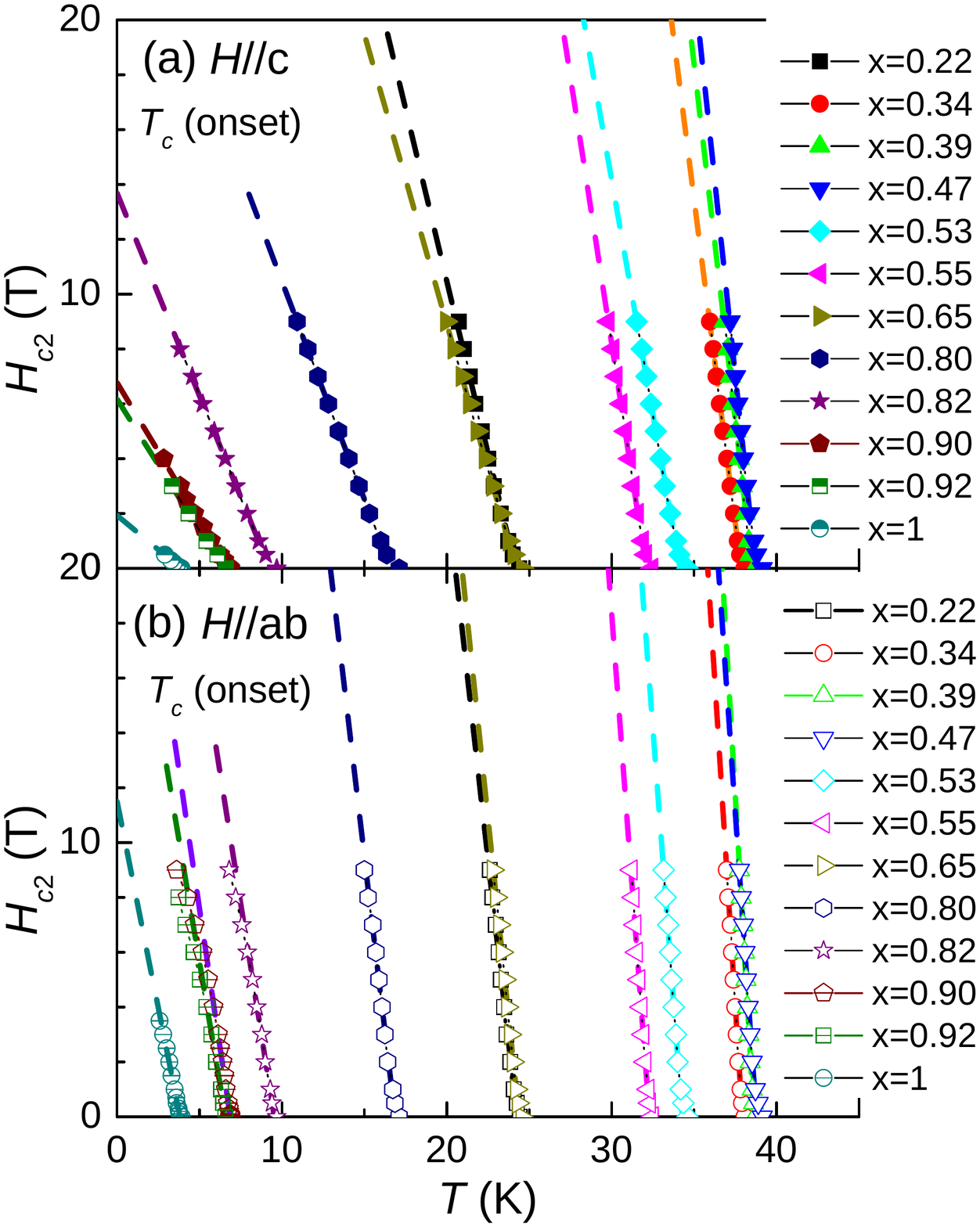}%
\caption{  (Color online) Summary of the $H_{c2}(T)$ curves, determined using onset criterion in temperature-dependent resistivity measurements, Figs.~\ref{Hc2rowunder},\ref{Hc2rowslight},\ref{Hc2rowheavy} for BaK12 single crystals $0.22 \leq x \leq 1.0$, in configurations $H \parallel c$ (top panel) and $H \parallel ab$ (bottom panel). Lines show linear fits of the data for fields close to zero-field $T_c(0)$ neglecting slight upturn in the lowest fields. The linear fits were used to determine slopes of the lines $dH_{c2}(T)/dT$ and evaluate zero-temperature $H_{c2}(0) =-0.70T_c(0) dH_{c2}/dT$, as shown in Fig.~\ref{Hc2summary} below.
}%
\label{Hc2slopes}%
\end{figure}

\begin{figure}[tbh]%
\centering
\includegraphics[width=7cm]{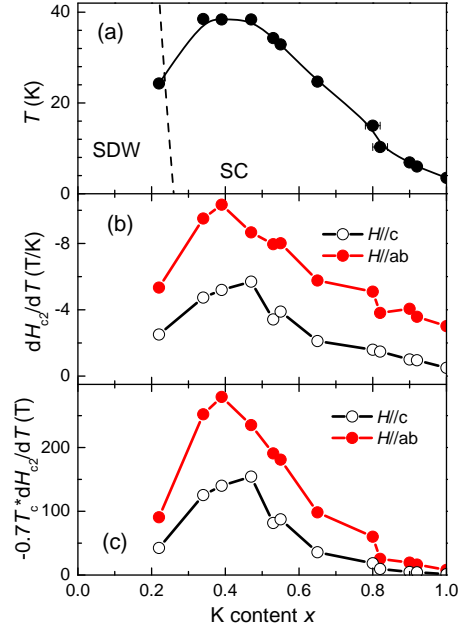}%
\caption{  (Color online) Doping evolution of the slope of $H_{c2}(T)$ curves close to zero-field $T_c(0)$, $dH_{c2}(T)/dT$ (panel b), and of the extrapolated $H_{c2}(0) =-0.70T_c(0) dH_{c2}/dT$ (c), shown in comparison with doping evolution of the superconducting transition temperature $T_c (x)$ (a). 
}%
\label{Hc2summary}%
\end{figure}

For the samples with $x$= 0.22, 0.34, 0.39, 0.47, 0.53, and 0.55, the $H_{c2}(T)$ curves show positive curvature close to $T_c(0)$ for lowest fields below $H$=1~T.  Going further below $T_c(0)$, the $H_{c2}(T) $ gets practically $T$-linear. This is exactly the range which we use for determination of the $dH_{c2}/dT$  slope (Fig.~\ref{Hc2slopes}) and evaluation of $H_{c2}(0)$ as $H_{c2}(0) =-0.70T_c(0) dH_{c2}/dT$ (as shown in Fig.~\ref{Hc2summary}). For the heavily overdoped samples $x$= 0.80, 0.82, 0.90, 0.92, and 1, the $H_{c2}(T)$ curves in configuration $H \parallel ab$ show a clear decrease of slope on cooling with a tendency to saturation, whereas for $H \parallel c$ the curves remain linear. The saturation in $H \parallel ab$ reflects paramagnetic Pauli limiting \cite{CC}. Similar saturation behavior is seen in underdoped samples \cite{Yuan}.


In Fig.~\ref{Hc2summary} we summarize the doping evolution of the slope of the temperature dependent upper critical field for field orientations along $c$-axis (open black circles) and along the plane (closed red circles) (middle panel). In the Werthamer-Helfand-Hohenberg (WHH) theory \cite{WHH} of the upper critical field for orbital limiting mechanism, $H_{c2}(0)=-0.7T_c(0) dH_{c2}/dT $. In the bottom panel of Fig.~\ref{Hc2summary} we plot $H_{c2}(0)$ estimated using WHH formula as $T_c(0) dH_{c2}/dT$. Note huge values of $H_{c2,c} >$100~T for compositions close to optimal doping. Interesting, the $H_{c2,c}(x)$ and especially $H_{c2,ab}(x)$ dependence, middle panel of Fig.~\ref{Hc2summary}, peaks at 0.39 and is much sharper than $T_c(x)$ dependence. 

\subsection{ Doping evolution of the anisotropy parameter $\gamma$}


\begin{figure}[tbh]%
\centering
\includegraphics[width=7cm]{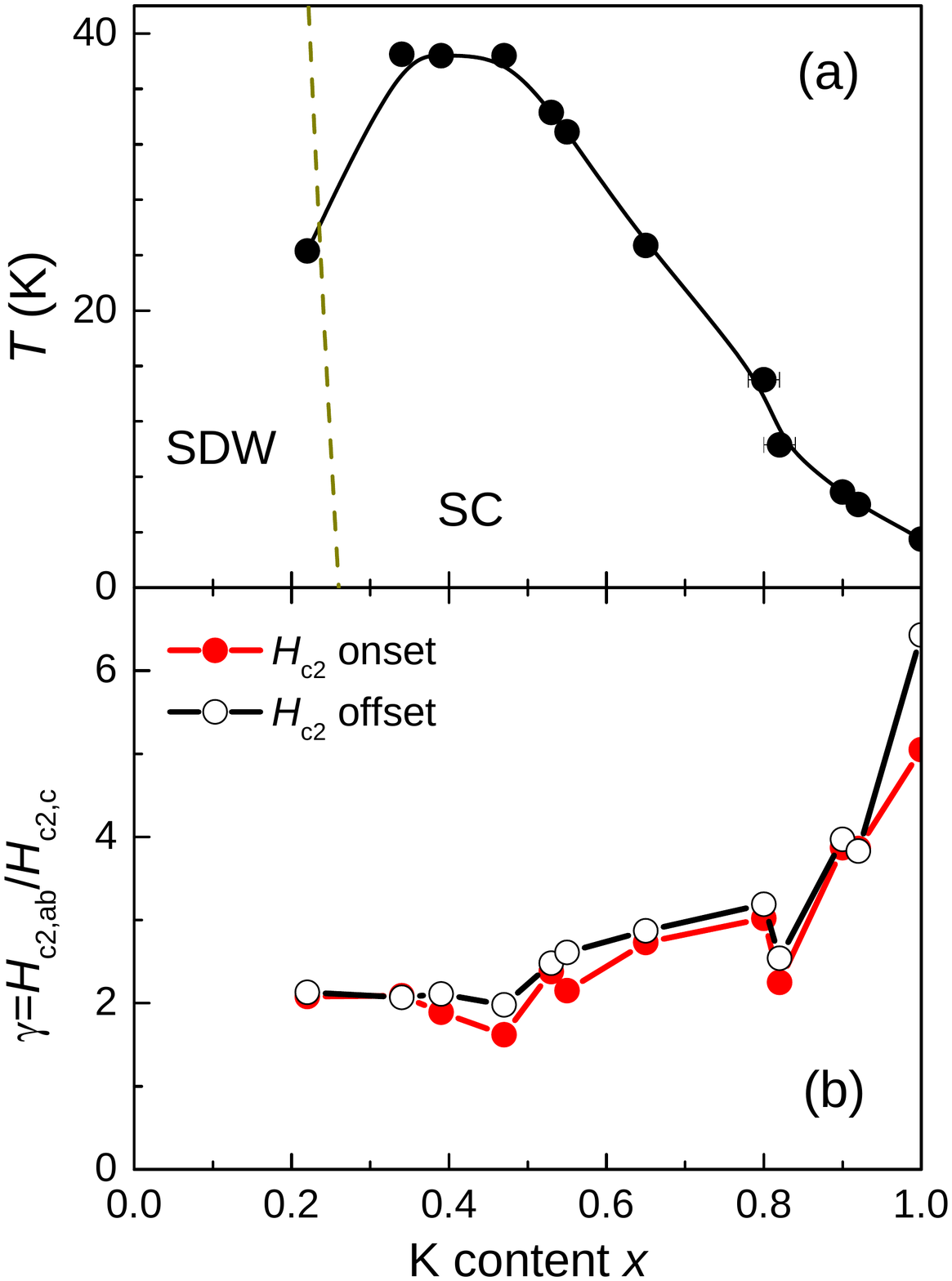}%
\caption{  (Color online) Doping evolution of anisotropy of the upper critical field $\gamma \equiv H_{c2,ab}/H_{c2,c}$ (bottom panel), shown in comparison with doping evolution of the superconducting transition temperature $T_c (x)$ (top panel). 
}%
\label{gamma}%
\end{figure}

In Fig.~\ref{gamma} we plot doping evolution of the anisotropy of the upper critical field $\gamma (x)$. It can be seen that $\gamma$ increases approximately two times, from 2 to 4 to 5 (depending on criterion) with increasing K doping levels. The increase starts in heavily overdoped compositions $x>$0.82, not far from the point where Fermi surface topology change was found in angle-resolved photoelectron spectroscopy (ARPES) studies \cite{ARPESFS} and where magnetism of the compounds changes according to neutron scattering \cite{Castellan,fluctuations} and NMR \cite{Kohori} studies. According to ARPES studies the electron sheet of the Fermi surface transforms to four tiny cylinders. Since electron sheets have largest contribution of $d_{z^2}$ orbitals, and are most warped, it is natural to expect anisotropy increase close to $x$=1 end of the doping phase diagram, in line with the upper critical anisotropy increase with $x$. 

Several previous studies of $H_{c2}$ anisotropy for selected $x$ close to optimal doping in BaK compounds were performed in high magnetic fields up to 60~T in samples with $T_c$=28.2~K ($x$=0.4)  \cite{Yuan}, $T_c$=32~K ($x$=0.45) \cite{NiNiK2}, and $T_c$=38.5 K ($x=$0.32) \cite{Gasparov}. They found anisotropy decreasing on cooling, which was presumably caused by contribution of paramagnetic effect for $H_{c2,ab}$. 

Similar to high-field studies in single crystals of other iron-based superconductors BaCo122 $x$=0.14 \cite{Kano,Gasparov}, NdFeAsO$_{0.7}$F$_{0.3}$  \cite{32Jaroszynski}, LiFeAs \cite{Cho}, and FeTe$_{0.6}$Se$_{0.4}$ \cite{34Khim,FeSeBalakirev}, we find rough linear increase of the $H_{c2,c}(T)$, but concave dependence with a tendency for saturation for $H_{c2,ab}$.  For all compounds of iron based superconductors the anisotropy ratio $\gamma$ at $T_c(0)$ is in the range 2 to 5, similar to our finding in BaK122, with Ca$_{10}$(Pt$_3$As$_8$)(( Fe$_{1-x}$Pt$_x$)$_2$As$_2$)$_5$ with $x=$0.09  \cite{NiNi1038Hc2}, SmFeAsO$_{0.85}$F$_{0.15}$ \cite{WelpSm} and LaFe$_{0.92}$Co$_{0.08}$AsO \cite{Balicas1111} being exceptions, with $\gamma \approx $7 to 8.

Additional contribution to the doping evolution of the anisotropy of the upper critical field can come from evolution of the superconducting gap structure \cite{KoganProzorov}. Initial high-resolution ARPES study on optimally doped samples with $x$=0.4 revealed a superconducting large gap ($\Delta \sim$12~meV) on the two small hole-like and electron-like Fermi surface sheets, and a small gap ($\sim$6~meV) on the large hole-like Fermi surface \cite{Ding}. In heavily overdoped KFe$_2$As$_2$, the Fermi surface around the Brillouin-zone center is qualitatively similar to that of composition with $x$=0.4, but the two electron pockets are absent due to an excess of the hole doping \cite{5Sato}. ARPES study over a wide doping range of BaK122 discovered that the gap size of the outer hole Fermi surface sheet around the Brillouin zone center shows an abrupt drop with overdoping (for $x \geq$0.6) while the gaps on the inner and middle sheets 
 roughly scale with $T_c$ \cite{6Malaeb}.

\subsection{Linear relation between $H_{c2}(T)$ slope and $T_c$}

\begin{figure}[tbh]%
\centering
\includegraphics[width=7cm]{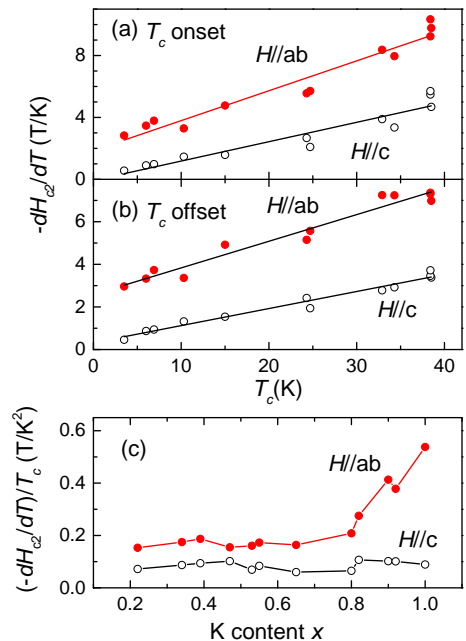}%
\caption{  (Color online) Dependence of the slope of the upper critical field, $dH_{c2}(T)/dT$ at $T_c$,  on the superconducting transition temperature $T_c$ for magnetic field parallel to tetragonal $c$-axis (open black circles) and parallel to the $ab$ plane (solid red circles) using onset (top panel) and offset (bottom panel) criteria.). 
}%
\label{linearslopeTc}%
\end{figure}

The high values of the critical fields in iron pnictides are determined by their short coherence lengths in 1 to 3 nm range \cite{Gurevich}, due to their high $T_c$ and low Fermi velocities, $v$, with $\xi \sim \hbar v/2 \pi k_B T_c$. 
Discussing the reasons for remarkable proportionality of the slopes of $dH_{c2}/dT$ to $T_c$ for $H\parallel c$ shown in Figs.~\ref{linearslopeTc}, we recall that  in {\it clean} isotropic s-wave materials,   
      \begin{equation}
 H_{c2} = -\frac{  \phi_0 (1-T/T_c)}{ 2\pi\xi_0^2}  \,, \quad \xi_0\sim\frac{\hbar v}{\Delta_0}\propto \frac{v}{T_c}  \,,  
\label{slope-clean}
\end{equation}

so that   the slope $H_{c2}^\prime \propto T_c$.  
For the dirty case $H_{c2}^\prime$ is $T_c$ independent; indeed,

      \begin{equation}
 H_{c2} \propto\frac{  1-T/T_c}{  \xi_0\ell}  \,, 
\label{slope-dirty}
\end{equation}
 where $\ell$ is the $T$ independent mean-free path.

We should mention that a strong pair breaking could be another reason for $dH_{c2}/dT\propto T_c$.  
For a gapless uniaxial material, the slope of the upper critical field along the $c$ direction near $T_c$ is given by\cite{K2009}

      \begin{equation}
\frac{dH_{c2,c}}{dT}  = -\frac{4\pi \phi_0 k_B^2 }{ 3\hbar^2\langle
\Omega^2 v_{ab}^2    \rangle}\,T_c  \,.  
\label{slope}
\end{equation}

Here ($\Omega(\bm k_f)$ describes the anisotropy of the order parameter and is assumed to have a zero Fermi surface average, $\langle
\Omega     \rangle=0$, which is the case for the d-wave or, approximately, for the $s^{\pm}$ symmetry). 

In our view, the first reason, i.e. the long mean-free path, is a probable cause for $dH_{c2}/dT\propto T_c$. Studies of thermal conductivity \cite{ReidSUST} and London penetration depth \cite{Prozorovreview} at optimal doping suggest full gap, which is inconsistent with the idea of gapless superconductivity. In Fig.~\ref{linearslopeTc} we verify linear relation for BaK122 over a broad doping (and as a consequence $T_c$) range, using onset (top panel a) and offset (middle panel b) criteria. The relation indeed holds very well, especially for $H \parallel c$ configuration where the $H_{c2,0}(x)$ curves extrapolate to zero on $T_C \to 0$. This suggests that there is no gross change in the Fermi velocity over the whole doping range. This quadratic $H_{c2}(T_c)$ relation is grossly violated in pure KFe$_2$As$_2$  in which the relation is linear \cite{YLiu}. 
For $H \parallel ab$ the $H_{c2,0}(x)$ curve is also close to linear, but does not extrapolate to zero on $T_c \to 0$. This deviation may be suggestive that Fermi velocity for transport along $c$-axis is strongly increasing in BaK compositions with lowest $T_c$ close to $x$=1.

Another way to check the linear relation between the slope of the upper critical field and $T_c$ is to plot their ratio, as shown in the bottom panel (c) of Fig.~\ref{linearslopeTc}. Plotting data this way reveals one difficult to recognize feature. The data for $H \parallel c$ indeed show constant and doping independent ratio $\frac {dH_{c2}/dT}{T_c}$. The ratio for $H \parallel ab$ remains constant for most of the phase diagram and then increases rapidly for $x>$0.8, showing that the increase of the anisotropy in this range is caused by decrease of the Fermi velocity, as one would expect for more anisotropic materials.

\section{Conclusions}

Using an inverted temperature gradient method we were able to grow large and high quality single crystals of (Ba$_{1-x}$K$_x$)Fe$_2$As$_2$ with doping range spanning from underdoped to heavily overdoped compositions ($0.22 \leq x \leq 1$). We show that high vapor pressure of K and As elements at the soaking temperature is an important factor in the growth of single crystals of BaK122. When setting the top zone as the cold zone, on cooling the nucleation starts from the surface layer of the liquid melt. It is also assisted by the vapor growth, because surface layer also saturates first due to the evaporation of K and As. The crystallization processes from the top of a liquid melt helps to expel impurity phases during, compared to the growth inside the flux. For the whole doping range $0.22 \leq x \leq 1$, we harvested large crystals with in-plane size up to 18$\times$10 mm$^2$. The crystals show very sharp superconducting transitions (less than 1~K) in dc magnetic susceptibil
 ity measurements for the optimal doping 0.34$\leq x \leq$0.55 and extremely overdoping 0.82$\leq x \leq$1 regimes. Relatively broad transitions are observed in the samples $x$=0.65 and 0.80, due to a broader distribution of Ba and K atoms and a tendency to K clustering in the lattice \cite{20Yeoh,Song}. 

In-plane electrical resistivity shows systematic evolution with doping. It perfectly follows $T^2$ dependence in the overdoped compositions with doping-independent slope over the range 0.80 to 1. Close to optimal doping the dependence deviates from pure $T^2$ functional form and can be described either as a sum of $T$-linear and $T^2$ contributions, similar to electron-doped materials \cite{NDL}, or using a power-law function with exponent $n \approx$1.5. 

The anisotropy of the upper critical field shows rapid change in the heavily overdoped regime, concomitant with Fermi surface reconstruction.
The slope of the $H_{c2}(T)$ curves scales with zero-field $T_c$ of the samples, suggesting nearly doping independent Fermi velocity.

\section{Acknowledgements}

This work was supported by the U.S. Department of Energy (DOE), Office of Science, Basic Energy Sciences, Materials Science and Engineering Division. The research was performed at the Ames Laboratory, which is operated for the U.S. DOE by Iowa State University under contract DE-AC02-07CH11358.


\end{document}